\documentclass[journal=jpccck,manuscript=article]{achemso}

\usepackage{graphicx}
\usepackage{amsmath}
\usepackage{amsfonts}
\usepackage{amssymb}
\usepackage{latexsym}
\usepackage[version=3]{mhchem}

\newcommand{\clithree}{\textrm{CLi}_{3}}
\newcommand{\clifour}{\textrm{CLi}_{4}}
\newcommand{\clifive}{\textrm{CLi}_{5}}
\newcommand{\clisix}{\textrm{CLi}_{6}}
\newcommand{\hyd}{\textrm{H}_{2}}

\newcommand{\oli}{\textrm{OLi}}
\newcommand{\olitwo}{\textrm{OLi}_{2}}
\newcommand{\olithree}{\textrm{OLi}_{3}}
\newcommand{\olifour}{\textrm{OLi}_{4}}

\author{S{\"u}leyman Er}
\affiliation{Computational Materials Science, Faculty of Science and Technology and MESA+
Research Institute, University of Twente, P.O. Box 217, 7500 AE Enschede, The Netherlands.}
\author{Gilles A. de Wijs}
\affiliation{Electronic Structure of Materials, Institute for
Molecules and Materials, Faculty of Science, Radboud University
Nijmegen, Heyendaalseweg~135, 6525~AJ Nijmegen,
The~Netherlands.}
\author{Geert Brocks}
\email{g.brocks@tnw.utwente.nl}
\affiliation{Computational Materials Science, Faculty of Science and Technology and MESA+
Research Institute, University of Twente, P.O. Box 217, 7500 AE Enschede, The Netherlands.}

\title{Hydrogen Storage by Polylithiated Molecules and Nanostructures}

\begin{document}

\date{\today}

\begin{abstract}
We study polylithiated molecules as building blocks for hydrogen storage materials, using first-principles calculations. $\clifour$ and $\olitwo$ bind 12 and 10 hydrogen molecules, respectively, with an average binding energy of 0.10 and 0.13 eV, leading to gravimetric densities of $37.8$ and $40.3$ weight $\%$ H. Bonding between Li and C or O is strongly polar and $\hyd$ molecules attach to the partially charged Li atoms without dissociating, which is favorable for (de)hydrogenation kinetics. CLi$_n$ and OLi$_m$ molecules can be chemically bonded to graphene sheets to hinder the aggregation of such molecules. In particular B or Be doped graphene strongly bind the molecules without seriously affecting the hydrogen binding energy. It still leads to a hydrogen storage capacity in the range 5-8.5 wt.$\%$ H.
\end{abstract}
\maketitle

\section{Introduction}
In recent years we observe a widespread effort aimed at finding alternatives to fossil fuels. The goal is to apply renewable resources that meet the increasing global demand for energy, while at the same time cut down CO$_{2}$ emissions \cite{pacala2004sws}. Hydrogen is a potential energy carrier, since it can be produced sustainably and is one of the most abundant elements \cite{coontz2004nss}. Burning hydrogen does not affect the environment as only water is produced. On board utilization of hydrogen on a large scale is hampered, however, by the lack of efficient hydrogen storage systems. Some of the methods suggested so far are high pressure gas tanks, low temperature liquid hydrogen tanks, chemical solid state storage such as metal-hydrides, and physisorption of hydrogen by materials with a high surface area \cite{schlapbach2001hsm}. Physisorption offers the possibility of storing hydrogen in molecular form. From kinetic considerations this is advantageous over chemical storage in atomic form, which requires dissociation of the hydrogen bond  and the formation of a hydride. Materials that are mainly based upon carbon are cheap, lightweight, and can have large internal surfaces. The bottleneck of physisorption by such materials lies in the weak interaction of hydrogen molecules with the host materials \cite{patchkovskii2005gnt}. As a result, very low temperatures or high pressures are necessary to attain meaningful storage capacities \cite{panella2005}. One can imagine encapsulating hydrogen in carbon nanocages, but the applicability of such systems is unknown at present \cite{pupysheva2008}. At the moment none of the above methods have achieved all requirements of a hydrogen storage system.

Previous studies have shown that isolated metal atoms or ions can bind hydrogen molecules, where the type of bonding depends on the metallic species. Transition metal atoms bind hydrogen molecules through weak multicenter chemical bonds called Kubas bonds \cite{niu1992bhm, gagliardi2004mha, durgun2006eth}. Alkali and alkaline earth ions bind hydrogen molecules through electrostatic (charge-multipole) and polarization (charge-induced multipole) interactions \cite{lochan2006}. Of particular interest is the lightweight Li$^{+}$ ion, which can bind up to six hydrogen molecules with an average binding energy of $0.19$ eV/H$_2$ \cite{chandrakumar2008c60}. This is close to optimal, as the binding energy should be around 0.15 eV/H$_2$ for storage of hydrogen gas at moderate pressures \cite{bhatia2006oca}. For this type of bonding it is essential that Li is ionized. Moreover, Li atoms need to be prevented from clustering, as, for instance, steric hindrance would result in a significant reduction of the amount of hydrogen molecules that can be bonded. Recent research therefore focuses on the hydrogen storage properties of Li-doped carbon nanostructures \cite{chandrakumar2008c60, chen1999hhu, zhou2004fps, deng2004nad, maresca2004ssp, zhu2004csh, cabria2005ehp, zhang2006cia, sun2006fps, fang2006ldp, cho2007hsl, han2007ldm, blomqvist2007mof5, chen2008fps}. A limited amount of Li can be introduced by doping without the Li atoms clustering, consequently restricting the wt.\% of adsorbed hydrogen significantly.

Polylithiated molecules are species that contain a large density of Li atoms. Such molecules have been studied experimentally and theoretically for a number of years, but never with an eye on their potential for hydrogen storage. The main focus has been on the interaction of single atoms such as C or O, with multiple Li atoms. The first polylithiated molecule, $\olithree$, has been identified by mass spectroscopy by Wu \textit{et al.} \cite{wu1979tpg} and its bonding characterized quantum chemically by Schleyer \textit{et al.} \cite{schleyer1982ehf}. The latter group has also predicted the existence of the carbon based polylithiated molecules $\clifive$ and $\clisix$ \cite{schleyer1983ehm}, which has later been confirmed experimentally by Kudo \cite{kudo1992ohc}. These studies have stimulated extensive research into the properties of CLi$_{n}$ and OLi$_{m}$ molecules \cite{rschleyer1996hum, lievens1999iph, jemmis1982lcg, ivanic1993nrs, kudo1993tcp, ponec2002ncl, zhizhong1998ais}. C$-$Li and O$-$Li bonds are strongly polar, resulting in a significant positive charge on the Li atoms. This should facilitate electrostatic bonding to hydrogen molecules, as discussed above.

In this paper we study the hydrogen storage properties of the polylithiated molecules CLi$_{n}$, $n = 3$-$5$ and OLi$_{m}$, $m = 1$-$4$. We find that each Li atom can bind multiple hydrogen molecules and that the binding energy increases with the partial charge on the Li atom. In particular, a $\clifour$ molecule binds up to $12$ and an $\olitwo$ molecule up to $10$ hydrogen molecules, with binding energies in the range of $0.10$-$0.14$ and $0.13$-$0.21$ eV/$\hyd$, respectively. The corresponding gravimetric storage densities are $37.8$ wt.\% H for $\clifour$H$_{24}$ and $40.3$ wt.\% H for $\olitwo$H$_{20}$. Such extremely high densities make polylithiated molecules fascinating building blocks for hydrogen storage materials. Aggregation of CLi$_{n}$ or OLi$_{m}$ molecules would be detrimental for the reversibility of hydrogenation. Therefore we explore the possibilities of attaching polylithiated molecules to carbon nanostructures, in particular graphene, in order to immobilize them. Using dopants such as B or Be, one can increase the binding of the molecules to the substrate, without weakening the binding of hydrogen to the molecules. This scenario leads to several interesting structures with gravimetric densities $>5$ wt.\% H.

\section{Computational Details}
The calculations are performed at the level of density functional theory (DFT), using the PW91 generalized gradient approximation (GGA) \cite{perdew1996gga}, and the projector augmented wave (PAW) method \cite{blochl1994paw, kresse1999upp}, as implemented in the Vienna \textit{ab initio} simulation package (VASP) \cite{kresse1993aim, kresse1996eis}. The PAW potentials treat the H 1\textit{s}, Li and Be 1\textit{s}, 2\textit{s}, B, C and O 2\textit{s}, 2\textit{p} as valence electrons. With a plane wave kinetic energy cutoff of $518$ eV the total energies are converged on a scale of $10^{-3}$ eV/atom. Geometries are optimized without any symmetry constraints, using the conjugate gradient (CG) algorithm. Convergence criteria of $10^{-4}$ eV and $0.02$ eV/\upshape{\AA} are applied for the energy difference between consecutive electronic steps and the total remaining forces on the atoms, respectively. Single molecules are treated within a cubic box with a cell parameter of $20$ \upshape{\AA}, using periodic boundary conditions and $\Gamma$-point sampling. In the graphene systems to be discussed below we use a 4$\times$4 in-plane graphene supercell and a 3$\times$3 regular \textbf{k}-point mesh to sample the Brillouin Zone (BZ). The final total energies and charge densities of the optimized structures are obtained using the corrected linear tetrahedron method\cite{blochl1994itm}.

\section{Results and Discussion}
\subsection{Polylithiated Molecules}
First we analyze the interaction of molecular hydrogen with CLi$_{n}$ and OLi$_{m}$ molecules. As the electron affinities of C and O are much larger than that of Li, the C$-$Li and O$-$Li bonds are very polar. If C and O become hypercoordinated by Li atoms, i.e. CLi$_{n}$ with $n\geq5$, or OLi$_{m}$ with $m\geq3$, the Li coordination shells become increasingly metallic and the average charge on the Li atoms decreases. The binding of hydrogen to these atoms is dominated by electrostatic and polarization interactions. Therefore, to maximize hydrogen binding energies, the charge on the Li atoms should be maximal and one needs to avoid hypercoordination of the C or O atoms. In $\clifour$ or $\olitwo$ all valence electrons are involved in C$-$Li and O$-$Li bonds, respectively, maximizing the charge on the Li atoms. The most stable geometries of $\clifour$ and $\olitwo$ have a high symmetry, i.e. $T_d$ and $D_{\infty h}$, respectively. All Li atoms in a molecule are then equivalent.

\begin{figure}[tb]
	\centering
		\includegraphics[width=8.5cm]{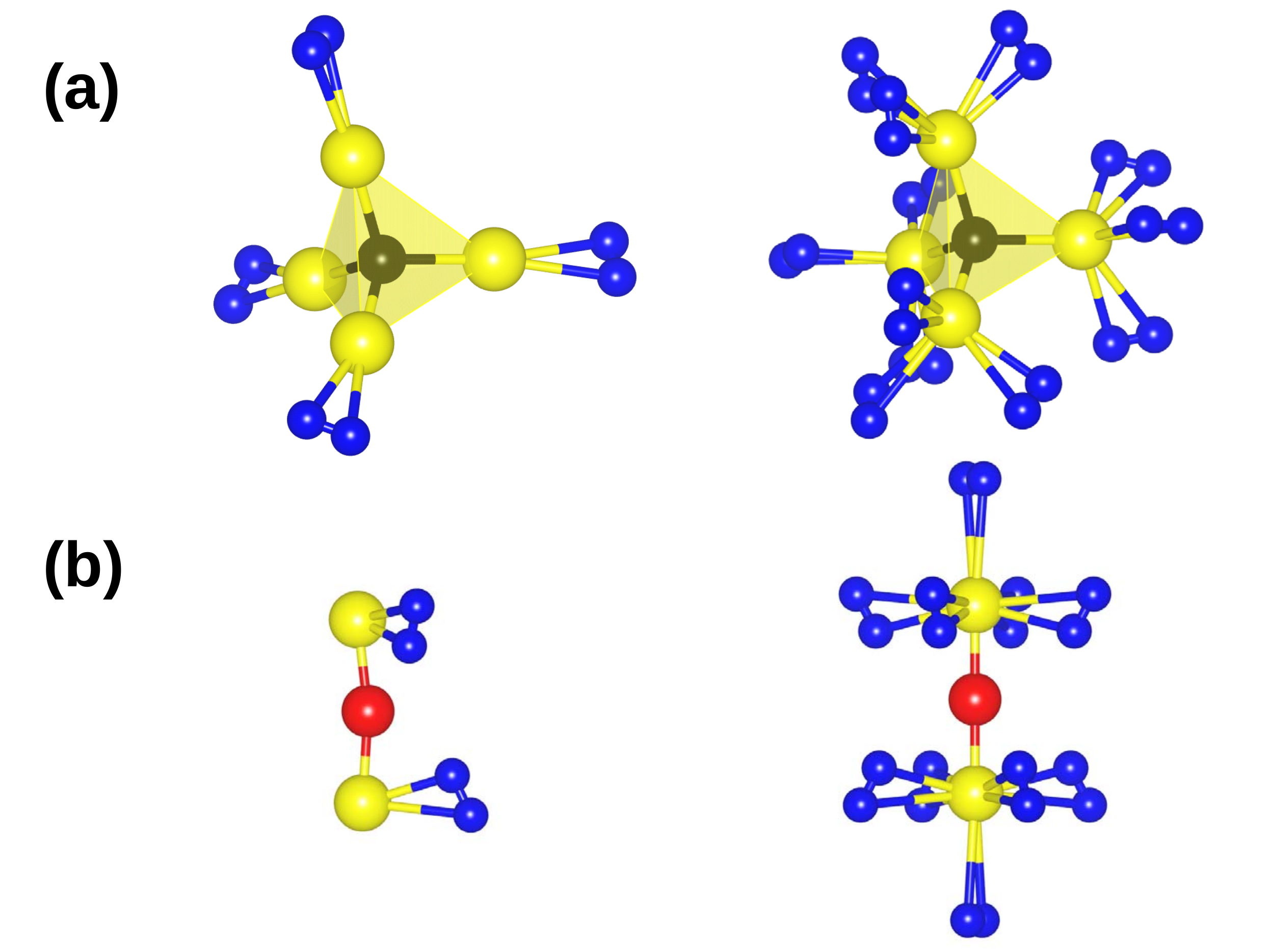}
		\caption{Optimized geometries of (a) CLi$_4$H$_8$, CLi$_4$H$_{24}$, and (b) OLi$_2$H$_4$, OLi$_4$H$_{20}$ complexes. Carbon, oxygen, lithium and hydrogen atoms are shown as black, red, yellow and blue spheres, respectively.}
	\label{molecules}
\end{figure}

A Bader charge analysis of the calculated self-consistent charge densities gives a charge of \texttt{+}$0.82e$ on each of the Li atoms in $\clifour$, and a charge of \texttt{+}$0.88e$ on the Li atoms in $\olitwo$ \cite{henkelman2006far}. These high charges facilitate the electrostatic bonding of hydrogen molecules to the Li atoms in these molecules, similar to that in isolated Li$^+$ ions \cite{chandrakumar2008c60}. Hydrogen molecules then cluster around the Li sites of the $\clifour$ and $\olitwo$ molecules. Examples of the structures of hydrogenated molecules are shown in \ref{molecules} and bond lengths before and after hydrogenation are given in \ref{table:CLi}.

\begin{table*}
\centering
\caption{Calculated bond lengths (\upshape{\AA}), Bader charges ($e$), hydrogen binding energies $E_{b}^{\hyd}$ (eV/$\hyd$) and gravimetric hydrogen densities of $\clifour$ and $\olitwo$ molecules. For Li$-$$\hyd$ the bonding distance is measured between Li atom and the center of H$-$H bonds. Average numbers are given and the numbers between () indicate the spreads.}
\label{table:CLi}
\begin{tabular}{lcccccc}
\hline
&\multicolumn{3}{c}{bond lengths (\upshape{\AA})} &       &       &       \\
\cline{2-4}
Molecule & C,O$-$Li & Li$-\hyd$ & H$-$H & charge on Li & $E_{b}^{\hyd}$ & H wt. \%  \\
\hline
$\clifour$        & $1.898(3)$ & $-$         & $-$        & $+0.819(2)$ & $-$     & $-$     \\
$\clifour$H$_{8}$ & $1.888(5)$ & $2.036(12)$ & $0.756(0)$ & $+0.827(2)$ & $0.14$ & $16.86$ \\
$\clifour$H$_{24}$& $1.908(3)$ & $2.158(7)$  & $0.763(2)$ & $+0.851(3)$ & $0.10$ & $37.82$ \\
$\olitwo$         & $1.643(0)$ & $-$         & $-$        & $+0.876(0)$ & $-$     & $-$     \\
$\olitwo$H$_{4}$  & $1.671(1)$ & $1.862(5)$  & $0.793(0)$ & $+0.869(0)$ & $0.21$ & $11.89$ \\
$\olitwo$H$_{20}$ & $1.711(1)$ & $2.140(144)$& $0.764(13)$& $+0.881(1)$ & $0.13$ & $40.29$ \\
\hline
\end{tabular}
\end{table*}

Also given in \ref{table:CLi} are the hydrogen binding energies, which are calculated by subtracting the total energy of the hydrogenated polylithiated molecules from the sum of the total energies of the isolated polylithiated molecules and the hydrogen molecules, divided by the number of hydrogen molecules. Each Li atom in a $\clifour$ molecule can bind three $\hyd$ molecules as shown in \ref{molecules}, which leads to 37.8 wt.\% H. The average binding energy decreases from $0.14$ eV for the first $\hyd$ molecule (per Li atom), to $0.10$ eV for the fully hydrogenated state. The decrease reflects a net repulsive interaction between the $\hyd$ molecules surrounding a Li atom. Steric hindrance prevents binding a fourth $\hyd$ (at least with a binding energy $>0.04$ eV). The $\hyd$ binding energies are consistent with those found for isolated Li$^{\texttt{+}}$ ions, bearing in mind that the Li atoms in CLi$_4$ have a somewhat reduced charge \cite{lochan2006}.

$\olitwo$ molecules bind hydrogen more strongly than $\clifour$ molecules. As the net charge on a Li atom in $\olitwo$ is larger than in $\clifour$ this can be expected if the binding is electrostatic. Because the Li atoms in $\olitwo$ are less crowded than in $\clifour$, each Li atom can accommodate up to five $\hyd$ molecules, before steric hindrance prevents binding of additional hydrogen. The average binding energy decreases from $0.21$ eV for the first $\hyd$ molecule (per Li atom) to $0.13$ eV in the fully hydrogenated state. It leads to $40.3$ wt.\% H for the OLi$_2$H$_{20}$ complex.

The fully hydrogenated CLi$_4$H$_{24}$ and OLi$_2$H$_{20}$ complexes keep a high symmetry, i.e. $T_d$ and $D_{2d}$, respectively. In intermediate hydrogenation stages the symmetry can be broken, see \ref{molecules}. After hydrogenation the C$-$Li and O$-$Li bonds are stretched by 1\% respectively 4\%. This is accompanied by a slight increase of the partial charge on Li by 4\% respectively 0.6\%. The optimized calculated bond length of an isolated H$_2$ molecule is 0.748 \AA. \ref{table:CLi} shows that in the complexes the H$-$H bond lengths increase by 1-6\%.  The limited size of these changes agrees with the notion of a weak interaction with the H$_2$ molecules. The absolute size of these changes generally correlates with the binding energy.

\begin{figure}[tb]
	\centering
		\includegraphics[width=8.5cm]{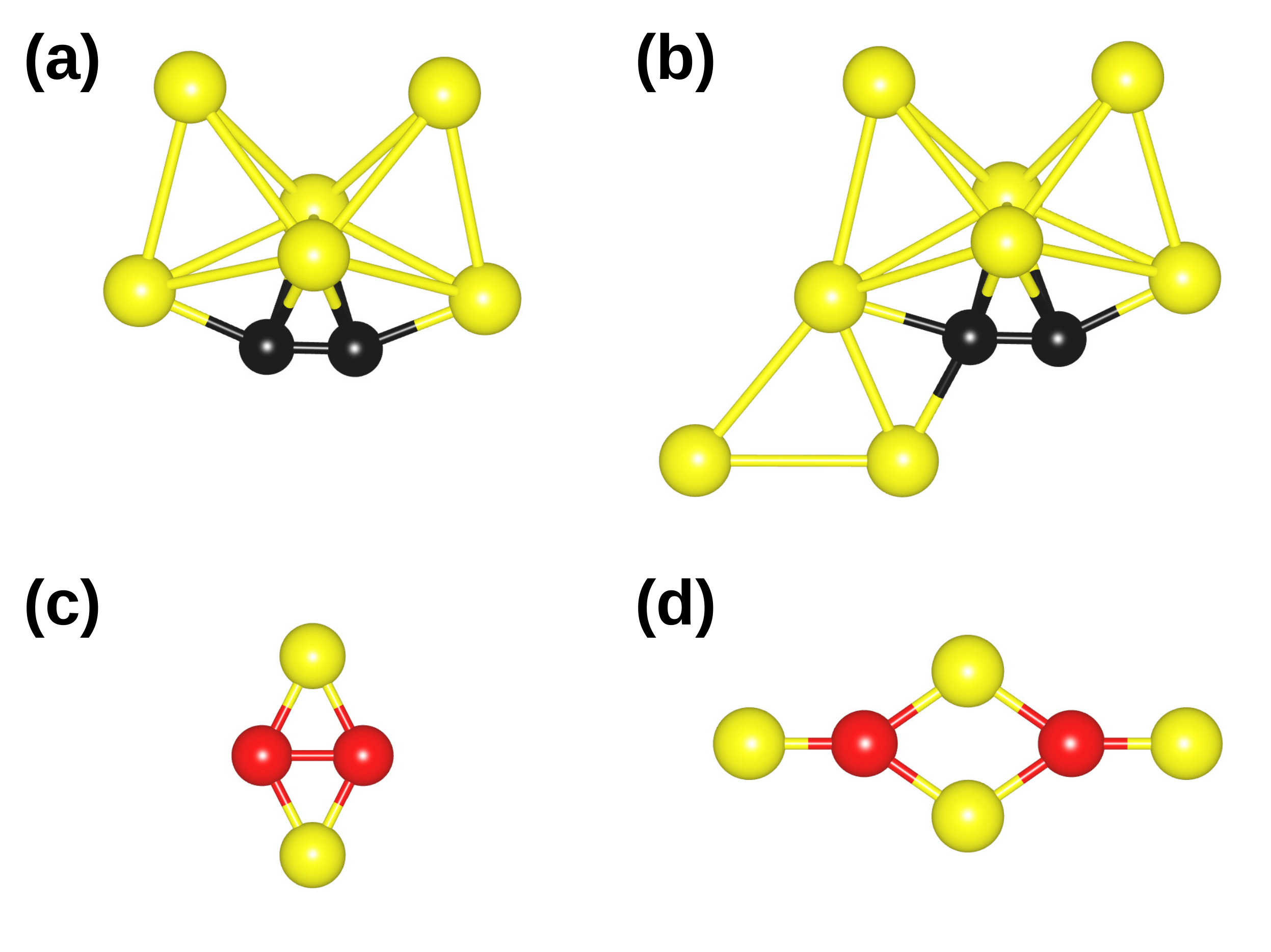}
		\caption{ Optimized geometries of (a) $(\clithree)_2$, (b) $(\clifour)_2$, (c) $(\oli)_2$ and (d) $(\olitwo)_2$.}
	\label{dimers}
\end{figure}

\subsection{Adsorption on Graphene}
It is clear that individual $\clifour$ and $\olitwo$ molecules can bind a substantial number of H$_2$ molecules. In order to maintain reversibility during repeated hydrogen loading and unloading cycles, one should however prevent the formation of clusters or aggregates of $\clifour$ or $\olitwo$ molecules. From a thermodynamic point of view such clusters could form. For instance, the equilibrium geometry of C$_2$Li$_8$ is shown in \ref{dimers}(b) \cite{ivanic1994isa}. The energy of this structure is $2.54$ eV/$\clifour$ lower in energy than that of two separate $\clifour$ molecules. Similarly, two $\olitwo$ molecules can form a cluster with a binding energy of $1.37$ eV/$\olitwo$, see \ref{dimers}(d). Presumably also structures of larger clusters exist that have a favorable binding energy. One could rely on kinetic barriers preventing clustering of polylithiated molecules. Here we consider alternative possibilities, which consist of immobilizing the molecules by binding them to a holder template. Microporous carbon and carbon nanostructures would be obvious substrates for these molecules.

As a prototype system we consider binding to a single sheet of graphite, i.e. to graphene. We use a 4$\times$4 graphene supercell, which contains $32$ atoms and has an optimized cell parameter of $9.866$ \AA. This cell is sufficiently large to prevent a direct interaction between molecules bonded to graphene in neighboring cells. The calculated binding energies of single $\clifour$ and $\olitwo$ molecules bonded to graphene are $0.73$ and $0.55$ eV, respectively. Both of these energies are considerably smaller than the corresponding molecular dimerization energies, indicating that the interaction with the substrate is not sufficiently strong. One has to consider increasing this interaction.

\begin{figure}[tb]
	\centering
		\includegraphics[width=8.5cm]{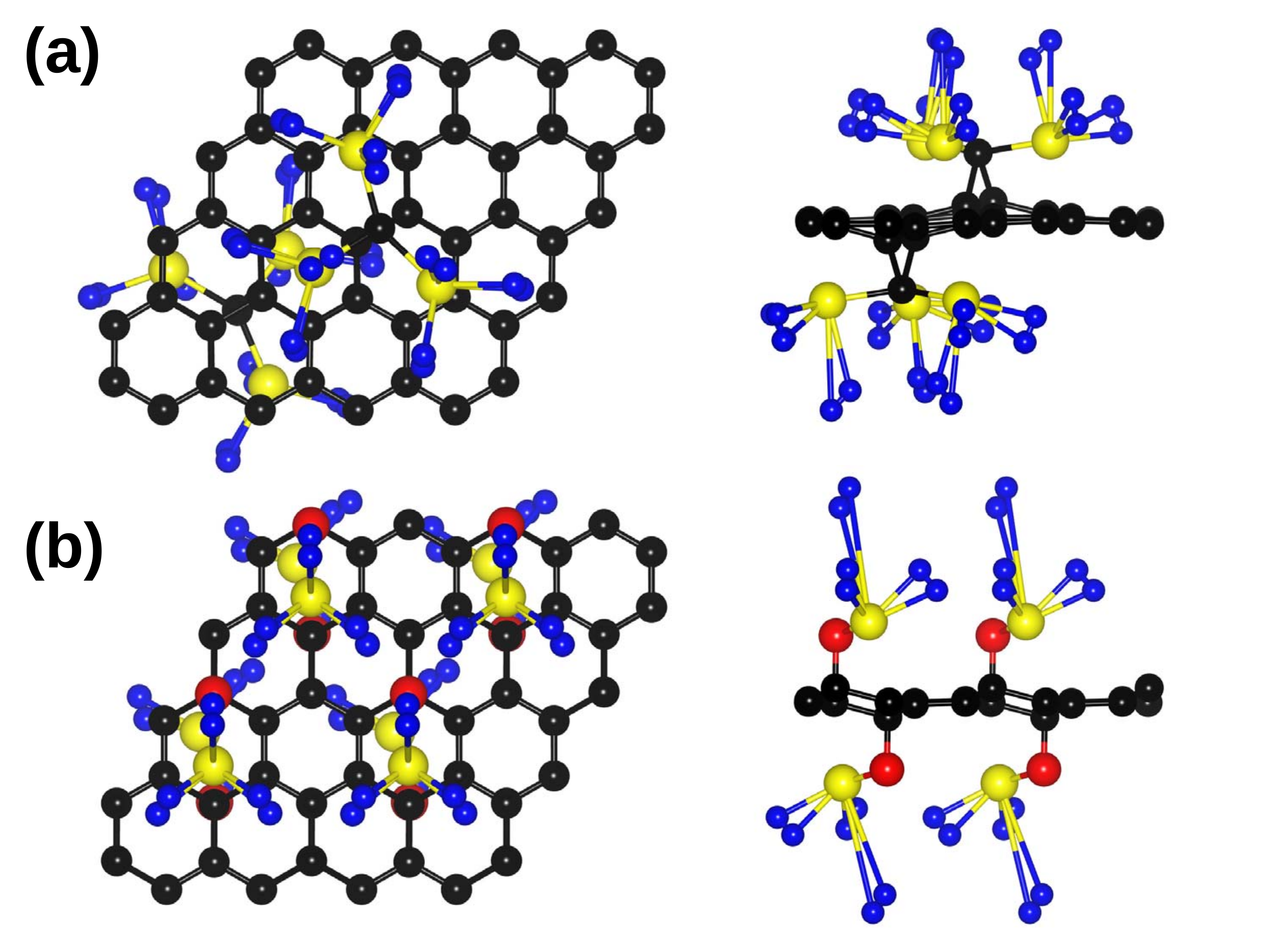}
		\caption{Top and side views of the optimized geometries of graphene binding (a) two $\clithree$ molecules, or (b) four $\oli$ molecules per supercell, in their fully hydrogenated forms.}
	\label{graphene_undoped}
\end{figure}

For this we pursue two different strategies consisting of modifying the molecules or the substrate, respectively. Removing one Li atom from $\clifour$ or $\olitwo$ strengthens the bonding to graphene. The calculated binding energy between $\clithree$ and graphene is $2.00$ eV. $\clithree$ molecules on graphene are preferably positioned over the center of C$-$C bonds. Both faces of graphene can be functionalized, which leads to a geometry as shown in \ref{graphene_undoped}(a). This structure is stable, but it is not the absolute minimum energy structure. C$_2$Li$_6$ in its equilibrium structure, cf. \ref{dimers}(a) \cite{sapse1995lco}, has a binding energy of $3.86$ eV/$\clithree$ with respect to two separate $\clithree$ molecules. It means that, although the $\clithree-$graphene structure is kinetically stable, there is still a thermodynamic driving force towards aggregation of $\clithree$ molecules.

$\oli$ molecules also bind strongly to a graphene substrate. Since an $\oli$ molecule is smaller than a $\clithree$ molecule, it is possible to quadruple the packing density of these molecules, as is shown in \ref{graphene_undoped}(b). $\oli$ molecules preferentially bind to the substrate through the formation of an O$-$C bond. The calculated binding energy is $2.22$ eV, which is comparable to the $2.26$ eV/$\oli$ aggregation energy of two $\oli$ molecules into O$_2$Li$_2$. The O$_2$Li$_2$ molecule has a planar structure as shown in \ref{dimers}(c). This means that the $\oli$-graphene structure is close to being thermodynamically stable.

\begin{figure}[tb]
	\centering
		\includegraphics[width=8.5cm]{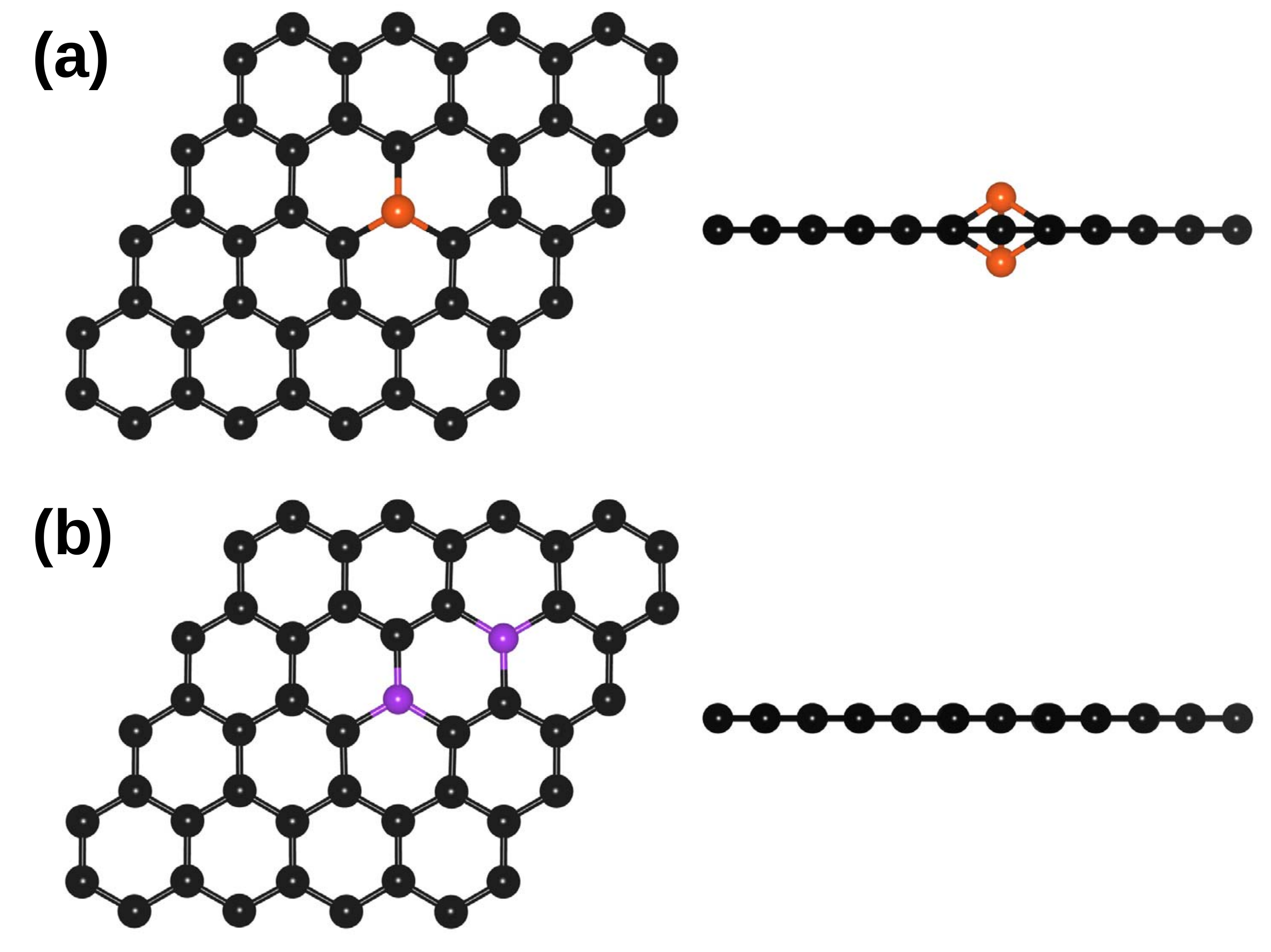}
		\caption{Top and side views of the optimized geometries of the (a) Be doped and (b) B doped graphene structures.}
	\label{doped}
\end{figure}

\begin{figure}[tb]
	\centering
		\includegraphics[width=8.5cm]{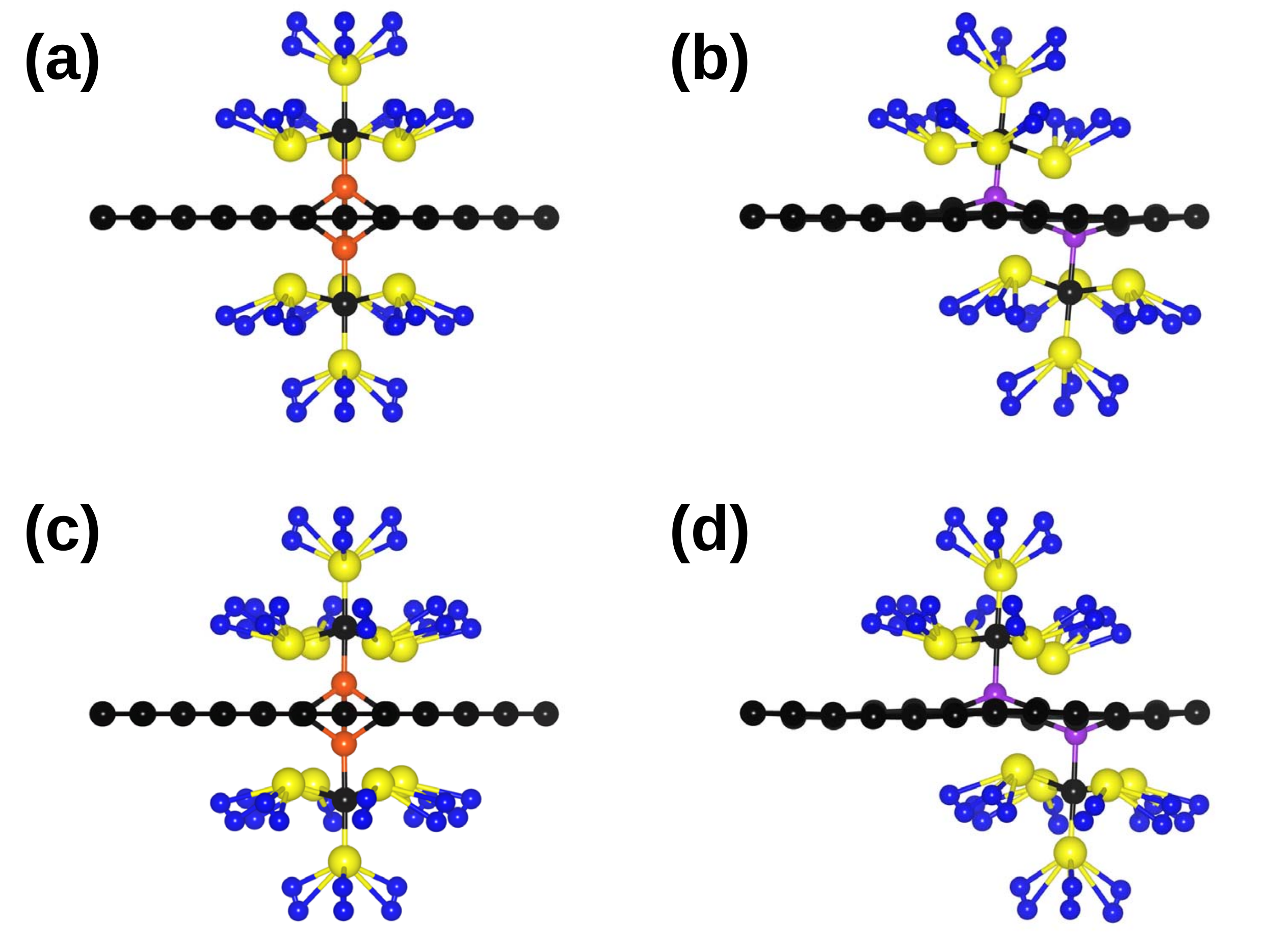}
		\caption{Side views of the fully hydrogenated structures of $\clifour$ on (a) Be doped and (b) B doped graphene and of $\clifive$ on (c) Be doped and (d) B doped graphene.}
	\label{doped_molecules}
\end{figure}

\subsection{Adsorption on Doped Graphene}
Another possible way of increasing the binding between polylithiated molecules and the substrate is by chemical modification of the graphene layer \cite{boukhvalov2008}. Introducing an impurity in graphene might for instance not only strengthen the interaction with CLi$_n$ and OLi$_m$ molecules, but could also maintain the positive charge on the Li atoms, and therefore the binding to $\hyd$ molecules. Here we consider modification of graphene by atomic substitution of C by B or Be atoms. Previous studies have shown that B doping of graphene does not cause any major deformation of the graphene structure \cite{endo2001stm}. B doping concentrations of as high as $17$ at. \% can be achieved in thin films \cite{hach1999ivd}. From experiment it has been suggested that the preferential geometry upon high B doping has two B atoms in para positions of an hexagonal ring in the graphene plane, as shown in \ref{doped}(b). Indeed, the stability of this geometry has recently been confirmed by a theoretical study \cite{miwa2008hab}. Here we consider such a double B substitution in our 4$\times$4 graphene supercell, which gives a doping concentration of $6.25$ at. \%.

As an alternative to B doping we also consider substituting C with Be atoms. Replacing one C with a Be atom does not alter the planar structure of graphene. It is energetically advantageous however to replace a single C atom by a pair of Be atoms. The two Be atoms are then displaced above and below the graphene plane by $0.855$ \upshape{\AA}, as shown in \ref{doped}(a), whereas the rest of graphene maintains its original structure. The Be atoms can act as pedestals that increase the distance between the graphene plane and the molecules we are going to attach, which could limit steric hindrance. All of the Be or B doped graphene structures have a nonmagnetic ground state.

Modifying graphene by Be or B doping increases its interaction with polylithiated molecules. The molecules preferentially bind to the dopant atoms and form C,O$-$B,Be bonds. From the point of view of binding strength, there is no need to modify the molecules by extracting a Li atom before attaching them to the substrate. $\clifour$ binds to a B or a Be site with binding energies of $2.91$, $3.81$ eV, respectively. The binding geometries are shown in \ref{doped_molecules}. The binding between $\olitwo$ and B or Be doped graphene has an energy of $2.49$, $3.76$ eV , respectively. Since all these binding energies are well above the molecular dimerization energies one expects that aggregation of these molecules is now suppressed.
Remarkably, doped graphene also forms strong bonds to other CLi$_n$ and OLi$_m$ molecules, including hypercoordinated ones such as $\clifive$, $\olithree$ and $\olifour$, leading to similar binding geometries. The binding energies are given in \ref{table:modgraphene} and examples of binding geometries are shown in \ref{doped_molecules}.

\begin{table*}[tb]
\centering
\caption{Binding energy (eV) between CLi$_n$ and OLi$_m$ molecules and (doped) graphene.}
\label{table:modgraphene}
	\begin{tabular}{lrrrrrrr}
\hline
    & $\clithree$&   $\clifour$&   $\clifive$&       $\oli$&   $\olitwo$&  $\olithree$&   $\olifour$\\
\hline
undoped         &      $2.00$&       $0.73$&             &       $2.22$&      $0.55$&             &             \\
Be doped        &      $4.12$&       $3.81$&       $3.96$&       $4.14$&      $3.76$&       $3.55$&       $3.00$ \\
B doped         &      $4.05$&       $2.91$&       $3.23$&       $3.45$&      $2.49$&       $3.13$&       $3.69$\\
\hline
\end{tabular}
\end{table*}

With the polylithiated molecules strongly bonded to (doped) graphene we now study the hydrogenation properties of these systems. For $\clithree$ and $\oli$ bonded to undoped graphene we find that each of the Li atoms in the molecules above and below the graphene plane can bind three $\hyd$ molecules, see \ref{table:hydrogen}. The average hydrogen binding energy is $0.10$ and $0.12$ eV, respectively, which is comparable to that in free $\clifour$ and $\olitwo$ molecules, compare \ref{table:CLi}. The resulting geometries are given in \ref{graphene_undoped}. The gravimetric densities in the $\clithree$- and $\oli$-graphene systems are $7.46$ and $7.85$ wt.$\%$ H, respectively.

As discussed above, doping of graphene with Be or B increases the binding of polylithiated molecules to the substrate. The effect of doping on the binding energy between hydrogen and these molecules is small in most cases, as can be observed in \ref{table:hydrogen}. Only for $\hyd$ on $\clithree$ and $\oli$ bonded to Be doped graphene the binding energy is significantly increased. The distance between the hydrogen molecules and the graphene plane is larger in these systems, as compared to the other cases. The gravimetric hydrogen density in the doped graphene systems is determined by the concentration of dopant atoms, since the polylithiated molecules are bonded to these atoms, as well as by the number of hydrogen molecules bonded to a single polylithiated molecule.

The latter is foremost determined by steric hindrance. A Li atom that is close to the graphene plane or to other Li atoms can in general capture only one or two $\hyd$ molecules, whereas the ones that are further away from the plane and other Li atoms can bind three $\hyd$ molecules. This is illustrated in \ref{doped_molecules}. We have fixed the doping concentration of graphene. In particular for the OLi$_m$ systems one can still increase the doping concentration without introducing too much steric hindrance, which would increase the gravimetric density substantially.

\begin{table*}[tb]
\centering
\caption{Average hydrogen binding energies $E_b^{\hyd}$ (eV/$\hyd$), number of hydrogen molecules \#$\hyd$ bonded per CLi$_n$ or OLi$_m$ species, and gravimetric hydrogen densities of polylithiated molecules bonded to (doped) graphene.}
\label{table:hydrogen}
\begin{tabular}{lccccccccc}
\hline
           & \multicolumn{3}{c}{undoped} & \multicolumn{3}{c}{Be doped} & \multicolumn{3}{c}{B doped} \\
           \cline{2-4} \cline{5-7} \cline{8-10}
           & $E_b^{\hyd}$ & \#$\hyd$ & H wt.\% & $E_b^{\hyd}$ & \#$\hyd$ & H wt.\% & $E_b^{\hyd}$ & \#$\hyd$ & H wt.\% \\
\hline
$\clithree$ & 0.10 & 9 & 7.46 & 0.15  & 7 & 5.83 & 0.11 & 8 & 6.72 \\
$\clifour$  &      &   &      & 0.09  & 9 & 7.17 & 0.09 & 9 & 7.29 \\
$\clifive$  &      &   &      & 0.08  & 11& 8.40 & 0.07 & 11& 8.53 \\
$\oli$      & 0.12 & 3 & 7.85 & 0.23  & 3 & 2.70 & 0.10 & 3 & 2.75 \\
$\olitwo$   &      &   &      & 0.10  & 6 & 5.10 & 0.09 & 6 & 5.19 \\
$\olithree$ &      &   &      & 0.12  & 6 & 4.96 & 0.10 & 6 & 5.04 \\
$\olifour$  &      &   &      & 0.08  & 9 & 7.06 & 0.06 & 9 & 7.18 \\
\hline
\end{tabular}
\end{table*}

Hydrogen binding energies decrease slightly for hypercoordinated polylithiated molecules, i.e. CLi$_n$, $n>3$ and OLi$_m$, $m>1$. \ref{doped_molecules} shows that this leads to a trade-off between hydrogen binding energy and gravimetric density. Nevertheless it is clear that there are several possible systems and structures that attain a gravimetric density higher than 5\% H with $\hyd$ binding energies in excess of 0.1 eV.

\section{Conclusions}
In conclusion, we have shown by means of first-principles DFT calculations that polylithiated molecules can be used as versatile building blocks for hydrogen storage materials. Li$-$C and Li$-$O bonds are strongly polar and the partially charged Li atoms can attract up to three to five $\hyd$ molecules with an average binding energy in the range 0.10-0.15 eV. $\clifour$ and $\olitwo$ molecules can then bind 12 and 10 hydrogen molecules, respectively, giving very high gravimetric densities of $37.8$ and $40.3$ weight $\%$ H. Binding polylithiated molecules to carbon nanostructures such as graphene can be used to immobilize them. In particular doping graphene with B or Be increases the binding energy. Although there is a trade-off between the size of the $\hyd$ adsorption energy and its gravimetric density in these systems, densities in the range 5-8.5 wt.\% H can be reached without seriously decreasing the adsorption energy.

\acknowledgement
This work is part of the research programs of Advanced Chemical Technologies for Sustainability (ACTS) and the Stichting voor Fundamenteel Onderzoek der Materie (FOM). The use of supercomputer facilities is sponsored by the Stichting Nationale Computerfaciliteiten (NCF). These institutions are financially supported by the Nederlandse Organisatie voor Wetenschappelijk Onderzoek (NWO).


\ifx\mcitethebibliography\mciteundefinedmacro
\PackageError{achemso.bst}{mciteplus.sty has not been loaded}
{This bibstyle requires the use of the mciteplus package.}\fi

\end{document}